\documentclass[conference]{IEEEtran}
 \pdfoutput=1
\usepackage{graphicx}
\usepackage{cite}
\usepackage{subcaption}
\usepackage{amsmath,amssymb,amsfonts}
\usepackage{algorithmic}
\usepackage{graphicx}
\usepackage{textcomp}
\usepackage{xcolor}
\usepackage{pifont}

\usepackage{etoolbox}
\makeatletter    
\patchcmd{\@makecaption}
  {\\}
  {.\ }
  {}
  {}
\usepackage{tikz}
\usepackage{xcolor}
\usepackage{soul}
\begin{document}
\newcommand{\proj}{\textit{Leaked-Web}}%{HPC-Stealing}}}
\newcommand*\circled[1]{\tikz[baseline=(char.base)]{
            \node[shape=circle,fill,inner sep=1.2pt] (char) {\textcolor{white}{#1}};}}
\title{{\proj}: Accurate and Efficient Machine Learning-Based Website Fingerprinting Attack through Hardware Performance Counters } % \vspace{-3ex}
%Stealing Webpage Fingerprint with low level hardware features}

% %%
% %% The "author" command and its associated commands are used to define
% %% the authors and their affiliations.
% %% Of note is the shared affiliation of the first two authors, and the
% %% "authornote" and "authornotemark" commands
% %% used to denote shared contribution to the research.

\author{Han Wang$^1$, Hossein Sayadi$^2$,  Avesta Sasan$^3$, Setareh Rafatirad$^3$, and Houman Homayoun$^1$\\

 $^1$University of California, Davis, CA, USA \\ 

$^2$California State University, Long Beach, CA, USA \\

 $^3$George Mason University, Fairfax, VA, USA 
 \\

 $^1$\{hjlwang, hhomayoun\}@ucdavis.edu, $^2$\{hossein.sayadi\}@csulb.edu,$^3$\{asasan, srafatir\}@gmu.edu

\vspace{-2ex}
}

% \affiliation{%

% \affiliation{%

% \affiliation{%

% \affiliation{%





\maketitle
\begin{abstract}
Users' website browsing history contains sensitive information, like health conditions, political interests, financial situations, etc. In order to cope with the potential website behavior leakage and enhance the browsing security, some defense mechanisms such as SSH tunnels and anonymity networks (e.g., Tor) have been proposed. Nevertheless, 
some recent studies have demonstrated the possibility of inferring website fingerprints based on important usage information such as traffic, cache usage, memory usage, CPU activity, power consumption, and hardware performance counters information. However, existing website fingerprinting attacks demand high sampling rate which causes high performance overheads and large network traffic, and/or they require launching an additional malicious website by the user which is not guaranteed. As a result, such drawbacks make the existing attacks more noticeable to users and corresponding fingerprinting detection mechanisms. In response, in this work we propose {\proj}, a 
novel accurate and efficient machine learning-based website fingerprinting attack through processor's Hardware Performance Counters (HPCs). 
{\proj} efficiently collects hardware performance counters in users' computer system at a significantly low granularity monitoring rate and sends the samples to the remote attack's server for further classification. {\proj} examines the
web browsers’ microarchitectural features using 
various advanced machine learning algorithms ranging from classical, boosting, deep learning, and time-series models. Our experimental results indicate that {\proj} based on a LogitBoost ML classifier using only the top 4 HPC features achieves 91\% classification accuracy outperforming the state-of-the-art attacks by nearly 5\%. Furthermore, our proposed attack obtains a negligible performance overhead (only $<$1\%) which is around 12\% lower than the existing hardware-assisted website fingerprinting attacks.    
%Compared to previous work, the proposed attack classification accuracy increases from 86.4\% to 91\% after employing the proposed {\proj}, and performance overhead is $<$1\% dropping from over 30\%. Compared to prior attacks, this work requires less samples, dropping from 15000 samples per second to 1 sample per second.
\end{abstract}
\vspace{1ex}

\vspace{-3ex}
\section{Introduction}
Over the last decades, the Internet has become an essential element of people's social lives to obtain new knowledge, information, and conduct businesses and daily tasks. However, new concerns about users' privacy have aroused from the transformation to the virtual world such that the users' browsing history could disclose some sensitive information about their background and motifs, such as financial status, sexual orientation, health conditions, or political views.  Therefore, through stealing users' online behaviors and access patterns, attackers could further induce personal and sensitive information of the users. %including sexual orientation or political beliefs and affiliations.
This has introduced web browser fingerprinting attacks to violate the privacy of the users by extracting and stealing the browsing history of the Internet users. To achieve this, a number of recent works  \cite{narayanan2012feasibility,rimmer2017automated} developed attacks that observe device-specific information and website access patterns such as packet sizes, packet timings, and direction of communication to infer the websites that the user is visiting with the aid of Machine Learning (ML) algorithms. 

% As a result, the main idea behind Web browser fingerprinting is capturing device-specific information and website access patterns of the users for purposes such as identification or security enhancement. 
% In contrast with
% other identification methods like cookies that rely on a unique identifier (ID) stored inside the browser, Web browser fingerprinting technique is stateless meaning that it it does not require to store the ID information inside the browser \cite{iqbal2020fingerprinting}.

To protect the online privacy of users, a number of researches %are dedicated 
have also proposed to hide %the all 
the network traffic of users \cite{torproject,luo2011httpos,booth2015not}. Tor network \cite{torproject} constructs an overlay network of collaborating servers, called relays. It encrypts the Internet traffic between users and web servers by transmitting the traffic between relays in a way that prevents external observers from identifying the traffic of specific users. The Tor Browser based on the Firefox web browser further protects users by disabling features that may be used for tracking users.% Though great progress made by such protection methodologies, website privacy is still 
%Though 
Despite considerable progress on developing users' privacy protections mechanisms, %provided, 
there are still 
a number of recent privacy violation attacks %which reply
that rely on the application of Machine Learning (ML) algorithms that are trained with %other 
computer systems' side-channel information collected when a website is open. These attacks stem from existing side-channel vulnerabilities %like 
like systems power analysis \cite{lifshits2018power}, CPU activity \cite{booth2015not}, on-chip cache memories \cite{shusterman2020website}, memory footprints \cite{jana2012memento}, storage \cite{kim2016inferring}, and hardware events \cite{gulmezoglu2017perfweb}. In this work, we comprehensively reviewed the existing studies on web browser fingerprinting attacks and identified some important challenges associated with these methods that could potentially result in underestimation of the security threats. 
% Such web privacy violation attacks obtain the data access pattern of the users %by sampling 
% at 1000-250000 per second sampling rate and employ Machine Learning (ML) classification techniques to identify %ing 
% the website %that is opened and 
% visited by the users with 80\% to 90\% accuracy. Though relatively effective, %for such attacks, 
% they require high data sampling rate which results in large data traffic, and high performance overhead. In addition, these attacks can be discovered easily. %What's more,
% Furthermore, high sampling rate increases the computational complexity and latency during testing phase while the website information may be used for following attacks where less latency is preferable, like Keydrown \cite{schwarz2018keydrown} with side-channel attacks. Furthermore, most side-channel information demands high level of isolation while multiple processes might run simultaneously in real-world, like CPU activity, memory. Lastly, most attacks \cite{shusterman2020website} adopt a malicious website where launch data monitoring, like the command-line version of wireshark \cite{rimmer2017automated} while visiting the malicious website undercut the threat of the attacks. However, new research \cite{iqbal2020fingerprinting} has shown that combining static and dynamic JavaScript analysis can detect such JavaScript-based attacks. 
\begin{table*}[!htb]
 \centering
     \caption{Recent Website Fingerprint attacks comparison and contributions of the {\proj} }
\scalebox{0.74}{
     \begin{tabular}{|c|p{2.5cm}|p{3cm}|p{1.7cm}|p{1.8cm}|p{1.5cm}|p{2.7cm}|p{2.7cm}|c|}
    \hline
    \textbf{Prior Works} & \textbf{Browser} & \textbf{Side-channel Information} & \textbf{Attack Model} & \textbf{Sampling Rate} & \textbf{Duration(s)} & \textbf{Performance Overhead} & \textbf{Machine Learning} & \textbf{Success Rate}\\\hline 
    Shane S et.al \cite{clark2013current}& Chrome &Power Consumption&Hardware&250,000&15&N/A&SVM&98\%\\\hline
Suman et.al \cite{jana2012memento}&Chrome, Firefox, Android & App memory footprint&Native Code&100,000&30-40&N/A&Customized Algorithm&N/A\\\hline
Hyungsub et.al\cite{kim2016inferring}  & Chromium Linux, Chrome&Quota Management API&JavaScript&N/A&~60&N/A&N/A&~90\%\\\hline
Pepe et.al\cite{vila2017loophole}  & Chromium Linux, Chrome Mac & Shared event loop&JavaScript&40000&5&N/A&Event Delay  Histograms, Dynamic Time Warping&76.7\%-91.1\% \\\hline
% \cite{cai2012touching}  & Firefox, Tor &  \\\hline
% \cite{cai2012touching}  & Firefox, Tor &  \\\hline
Anatoly et.al\cite{shusterman2020website}  & Firefox, Chrome, Safari, Tor Browser&Cache occupancy&JavaScript&500&30 &N/A&Deep Learning&82\%\\\hline
Berk et.al\cite{gulmezoglu2017perfweb}  & Firefox, Tor &Hardware Performance Counters&Native Code&25,000&5 & N/A&Classic, Deep Learning&86.4\%\\\hline
Sangho \cite{lee2014stealing}&Chromium,Firefox& GPU memory&Native& &N/A&N/A&Pixel sequence and histogram matching& 95.4\% \\\hline
% Power Usage \cite{clark2013current}&Chrome& Mac OS, Windows, Linux&Hardware&AC Power consumption &7.8k-250k \\\hline
Qing et.al \cite{yang2016inferring}& Chrome&Power usage & Hardware &10&200k&&Random Forest& $>$90\% \\\hline
\textbf{This Work ({\proj})}&Firefox&Hardware Performance Counters&Native Code&1&5-60&$<$1\%&Classic, Boosting, Deep Learning, Time-series&80\%-91\%\\\hline
    \end{tabular}}
    % \vspace{-5ex}
    %\hl{ in the table add the author name in the first column "ABC et al.,"}
    \label{tbl:contribution}
\end{table*}

\vspace{0.5ex}
\noindent
\textbf{Challenge \circled{1} High Sampling Rate and Performance Overheads:} %{Challenge 1: }}
Existing web privacy violation attacks obtain the data access pattern of the users %by sampling 
at 1000-250000 per second sampling rate and employ machine learning classification techniques to identify %ing 
the website's characteristics %that is opened and 
visited by the users with 80\%-90\% accuracy. Though relatively effective in terms of detection accuracy, %for such attacks, 
such attacks require significant data sampling rate which results in large data network traffic and high performance overhead. Therefore, these attacks are easily noticeable by users and/or the detection systems. 

% \noindent
% \textbf{\circled{2} {Challenge 2:}} 
% %In addition, 
% The high sampling rate causes obvious performance overhead that could be noticed by users or detection mechanisms. %like \cite{gulmezoglu2017perfweb} incurring 12\% shown in Figure \ref{fig:overhead}. 
% In addition, the number of samples captured per website for classification %visit
% ranging from 15,000 \cite{} to 4,000,000 which results in large network traffic and computational latency during the testing phase of the applied ML algorithm. As a result, these attacks could be discovered by the detection system. 
% These attacks can be discovered easily due to the high performance overhead caused by high sampling rate. And 
% Furthermore, high sampling rate increases the computational complexity and latency during the testing phase of the applied ML algorithm while the website information may be used for %following
% attacks where less latency is preferable, like Keydrown \cite{schwarz2018keydrown} with side-channel attacks. 
\vspace{0.5ex}
\noindent
\textbf{Challenge \circled{2} Limitation in Monitoring of Websites' Information:} %{Challenge 2:}} 
%Furthermore, 
%Most 
Recent website fingerprint attacks \cite{jana2012memento,kim2016inferring} adopt side-channel information such as CPU activity, memory usage, and storage to learn the browsing history of the users. %while 
Accurate collection of such information demands high level of isolation in which only a single website should be open on the system. Nevertheless, in real world scenarios a browser could open multiple websites at the same time which has been ignored in existing attack models.

\vspace{0.5ex}
\noindent
\textbf{Challenge \circled{3} The Need to Malicious Website Trigger:} %{Challenge 3: }}
Majority of privacy violation attacks \cite{shusterman2020website,vila2017loophole,kim2016inferring} adopt a malicious website %where 
to launch the data monitoring process (e.g., the command-line version of wireshark in \cite{rimmer2017automated}). However, visiting the malicious website by the user %undercut 
could lower the threat of the attacks. Moreover, new research studies such as \cite{iqbal2020fingerprinting} has shown that such attacks could be mitigated by combining static and dynamic JavaScript analysis that could successfully detect JavaScript-based attacks. 

\vspace{0.5ex}
\noindent
\textbf{Challenge \circled{4} Lack of Analysis on Monitoring and Number of Features:} %{Challenge 4: }
Our study indicates that prior website fingerprinting attacks have ignored to conduct a comprehensive analysis on the impact of different monitoring duration and number of features collected from website to infer accurate results. %of website inferring accuracy when monitoring duration and number features change. 
Nonetheless, such analysis is critical to thoroughly evaluate the effectiveness of the deployed attack threats and highlight the importance of adapting better protection mechanisms against such privacy violation attacks. 

% This work proposes an alternative website privacy violation attack which monitors hardware performance counters (HPCs) with low sampling rate and incurs trivial performance overhead. Furthermore, customized machine learning is adopted to further boost accuracy to XX\%, increased by XX\% compared to prior work. 

To address the above-discussed challenges, in this paper we thoroughly investigate the security threats brought by exploiting hardware-related features collected from processor's Hardware Performance Counters (HPCs) and machine learning classification techniques. The HPCs are special-purpose registers implemented into modern microprocessors to capture the trace of hardware-related events \cite{demme2013feasibility}. To this aim, we propose {\proj} a novel accurate and efficient machine learning-based website fingerprint attack model that exploits the information from performance counters with significantly low number of samples and performance overheads. Unlike prior works, the proposed {\proj} attack offers the lowest sampling rate which incurs the least performance overhead to the system.  {\proj} adopts advanced machine learning algorithms to further acquire website browsing history violating the privacy of the user. Since the HPC events in {\proj} could be collected from user space with no privileged access and no hardware overhead/modification, which makes the proposed fingerprinting more practical and efficient. To thoroughly analyze the effectiveness of {\proj}, %In this work,
we also investigate the impact of monitoring %interval
granularity and the number of required samples %required
on systems' performance overhead and attack success rate. %The %As shown in
%Table \ref{tbl:contribution} characterizes the proposed HPC-based {\proj} attack as compared with state-of-the-art website fingerprint attacks in terms of success rate, performance overhead, the number of samples. %It is safe to say that this work proposes a more powerful, less noticeable attack.

% The main contribution of this work can 

% We first demonstrate the threat posed by the attack that observes Hardware Performance Events and steal browsing history in terms of monitoring granularity, prediction accuracy. Compared to prior work \cite{}, the most prominent HPCs are investigated since the number of HPCs can be collected simultaneously is limited, ranging from 2 to 8. Secondly,  time-series classification and LSTM \cite{} are adopted to further empower the capability of the attacks. 

%In Table \ref{tbl:contribution}, the differences of prior work and the %attack prototyped proposed {\proj} attack in this work are summarized.
% \section{Characterization of Existing Works}
In Table \ref{tbl:contribution}, we have comprehensively analyzed the characteristics of the proposed HPC-based {\proj} attack as compared with state-of-the-art website fingerprinting attacks across various metrics such as target browser, attack model, performance overhead, the number of samples, deployed ML models, success rate, etc. It can be found that hardware-based, native code-based, and JavaScript-based are three main methods to exploit side-channel information and infer browsing history. However, JavaScript-based attacks \cite{shusterman2020website,vila2017loophole} can only be launched when the malicious website is visited, greatly undermining the capability of the attacks. Some other works \cite{kim2016inferring} also adopt native code attack model, while the memory or storage information can only be measured per browser instead of per website. Another drawback of previous works is the high sampling rate which can cause high performance overhead. For example, our analysis shows that monitoring hardware performance counters with 10,000 sampling rate could incur around 12\% performance overhead (will be discussed in detail in Section 3). %in Figure \ref{fig:overhead}. 
%Such high overhead makes attacks more noticeable for users or other detection software. 
Compared to previous work, this work achieves a high accuracy with lowest sampling rate and performance overhead. Furthermore, it does not require visiting malicious website to launch the attack, which offers higher flexibility and less restriction for {\proj}. %More specifically,
In particular, the main contributions of our work are summarized below:
%\vspace{-1ex}
\begin{itemize}
% \vspace{-0.5ex}
    \item We %evaluate
    examine the impact of various HPC features for the hardware-based website fingerprinting attack and identify the most prominent low-level features that are crucial to be protected from user space.
    \item The influence of the number of application traces (sampling rate) and samples per traces are investigated to evaluate the effectiveness and stealing capability of the attack when fewer traces and samples are available.
    \item We explore the impact of features monitoring on the overall performance (e.g., execution time) of the system. %and minimize it to only 1\% performance overhead. 
    Hence, our proposed privacy violation attack leads to the least sampling rate, performance overhead, and traffic as compared to the state-of-the-art attacks.
   % \item We demonstrate an attack with the least sampling rate, meaning it incurs the least performance overhead, traffic transferring.
    \item Various machine learning techniques are comprehensively implemented and evaluated to further demonstrate the %power
    effectiveness of the proposed HPC-based website attack, with %240 samples and
    91\% success rate while obtaining less than 1\% performance overhead.
\end{itemize}
% \vspace{-0.5ex}
% The remainder of this paper is organized as follows. The background and motivations are described in Section \ref{sec:bg}. Section \ref{sec:model} demonstrates design and procedure of implementing the {\proj}. Next,
% Section \ref{sec:eval} compares the HPCs-based detectors in terms of classification accuracy with various machine learning methods and different samples and traces. Lastly, Section \ref{sec:con} presents the conclusion of this study.

\section{Background and Related Works}
\label{sec:bg}

\begin{figure*}[!htb]
  \centering
\begin{subfigure}[b]{0.325\textwidth}
   \includegraphics[width=0.95\textwidth]{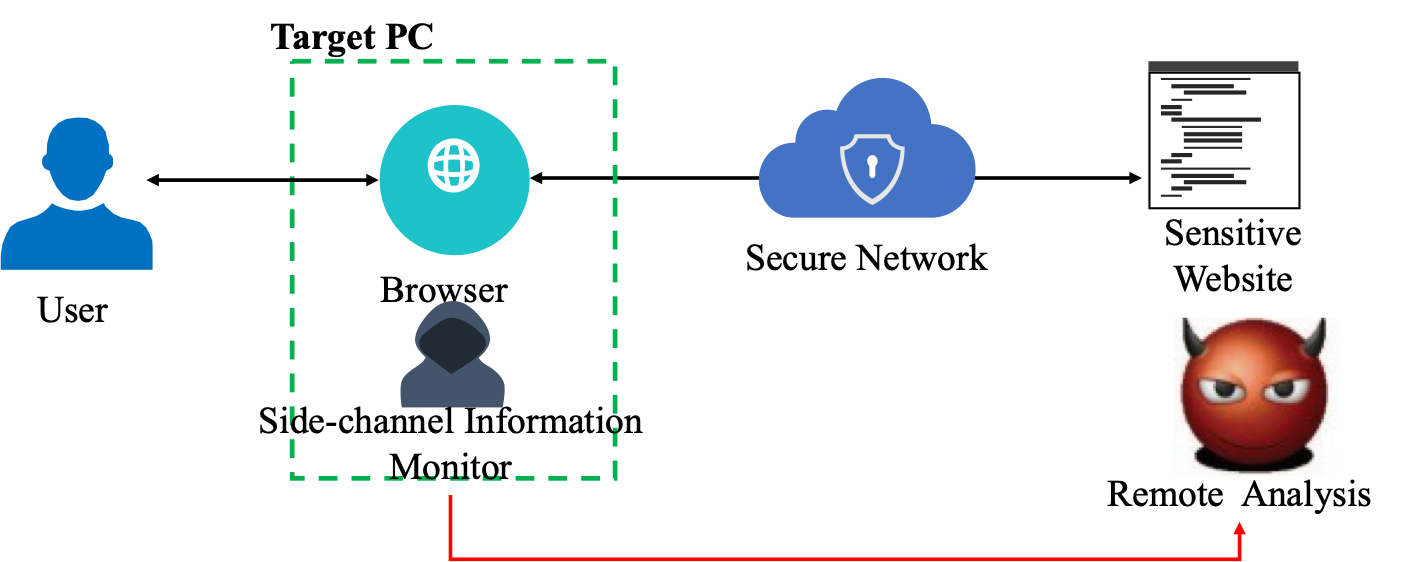}
  \caption{Native attack model}
     \label{fig:nativemodel}
\end{subfigure}
% \hspace{-2ex}
 \begin{subfigure}[b]{0.325\textwidth}
   \includegraphics[width=0.95\textwidth]{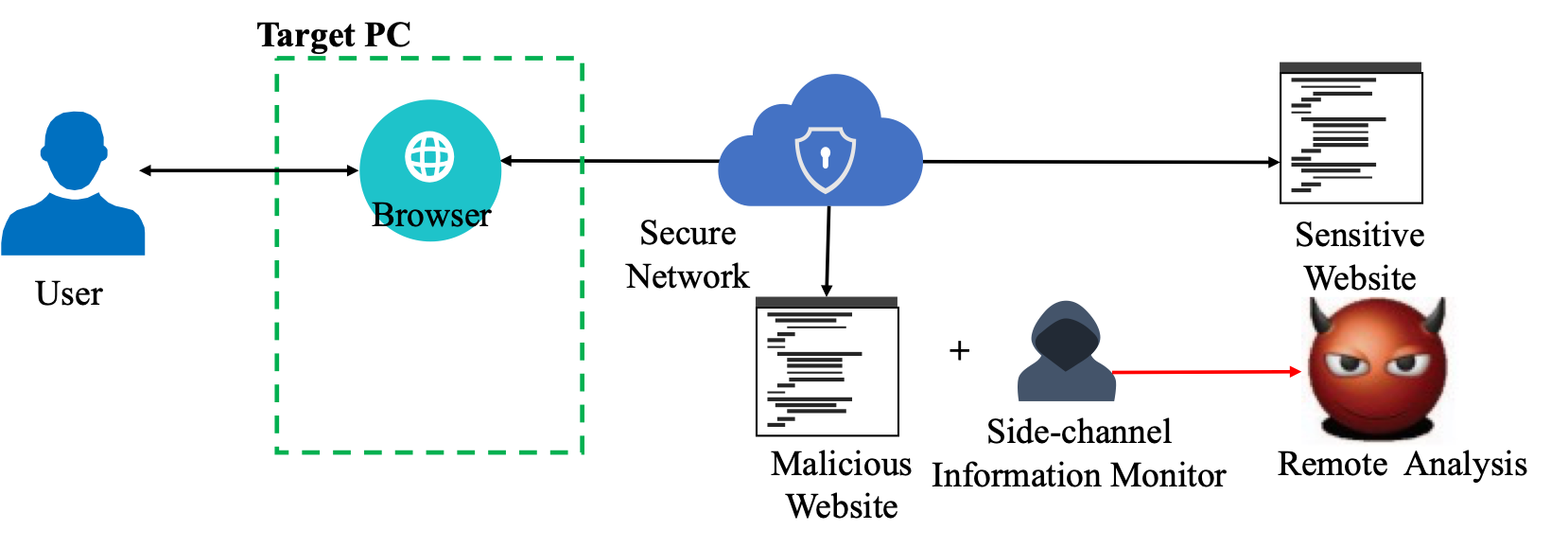}
     \caption{Malicious website attack model}
     \label{fig:remotemodel}
\end{subfigure} 
 \begin{subfigure}[b]{0.33\textwidth}
   \includegraphics[width=0.95\textwidth]{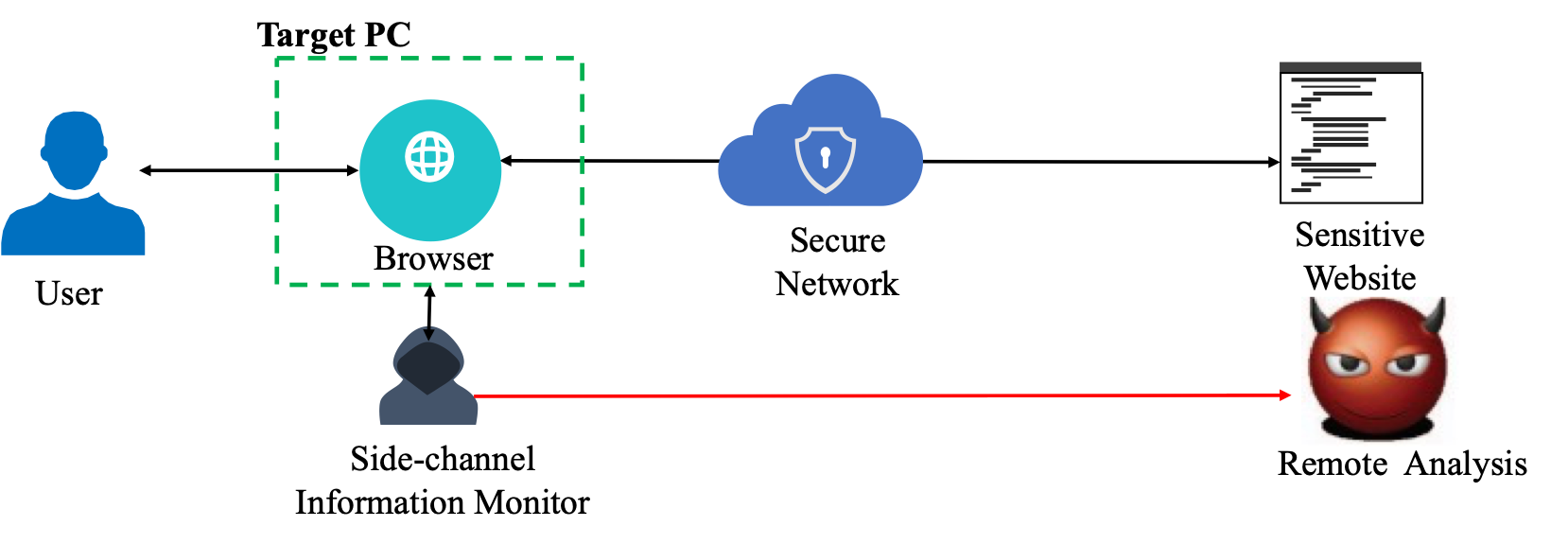}
  \caption{Hardware attack model}
     \label{fig:hardwaremodel}
\end{subfigure}
\caption{Website fingerprinting attacks threat model}
\end{figure*}

\subsection{Website Fingerprinting Attacks} 
Since website browsing history contains much sensitive information like medical status, political interest, etc., it is critical to prevent leakage, protect users' privacy, and enhance the system's security against potential cyber-attacks. However, prior research has demonstrated that fingerprinting attacks can be independent of operating systems and browsers and rely on side-channel information collected passively. Such attacks can be launched both remotely and locally, or via a peripheral device. The side-channel information exists across different computing abstracts, including hardware, system, and network. Once the side-channel information is collected, ML-based classification is leveraged to infer users' visited website information. %which websites users visit.
There are three popular threat models targeting stealing users' browsing history, including native attack model \cite{gulmezoglu2017perfweb}, malicious website attack model \cite{shusterman2020website}, and hardware attack model \cite{clark2013current}.

% \begin{figure}[!htb]
%   \centering
%   \includegraphics[width=0.45\textwidth]{pic/nativemodel.png}
%   \caption{Native attack model}
%     \vspace{-2ex}
%   \label{fig:remotemodel}
%     % \vspace{-2ex}
% \end{figure}
% \begin{figure}[!htb]
%   \centering
%   \includegraphics[width=0.45\textwidth]{pic/remotemodel.png}
%   \caption{Malicious website attack model}
%     \vspace{-2ex}
%   \label{fig:remotemodel}
%     % \vspace{-2ex}
% \end{figure}
% \begin{figure}[!htb]
%   \centering
%   \includegraphics[width=0.45\textwidth]{pic/hardwaremodel.png}
%   \caption{Hardware attack model}
%     \vspace{-2ex}
%   \label{fig:hardwaremodel}
%     \vspace{-2ex}
% \end{figure}

\subsubsection{Native Attack Model} In this attack model, the assumption is that the malicious code is resided in the host machine already, which can be done by inserting it into benign applications or downloading accidentally. With this model, attacks can be activated as long as the malicious codes are installed already and do not need users to open certain websites.

\subsubsection{Malicious Website Attack Model} 
By comparison, this attack model assumes that users click on a malicious website link; thereby, malicious JavaScript codes are executed in local computers. Compared to the native model, this model is more flexible to update the attack and can be less visible since local malware scan cannot detect malicious codes. For both models, the malicious codes are only in charge of side-channel information monitoring without compromising systems. 

\subsubsection{Hardware Attack model} 
As shown in Figure \ref{fig:hardwaremodel}, this model collects hardware properties, mainly power consumption, when various websites are opened. Some of them infer the side-channel information based on hardware monitoring components such as USB and power meter. Though physical access is needed under this mode, \cite{clark2013current} measures the power consumption and achieves 98\% website classification accuracy, posing a significant security threat to computer systems and privacy. 

%\hl{add more description here Jane: added}
% Taking fingerprinting attacks that exploit hardware performance counters (HPCs) as an example, attacks can profile the HPCs data for each website offline and use them to build a machine learning-based classifier model. And then attackers activate the HPCs monitoring in users' computers and send them to remote classification models that can output the website users are visiting, like "google.com". The research in \cite{perfwebpapge} demonstrated 86.3\% classification accuracy based on HPCs information.

% \begin{table}[!htb]
%   \centering
%     \caption{Perf Paranoid Setting \cite{linuxman}}
%   \scalebox{1}{
%   \begin{tabular}{|p{0.15\textwidth}|p{0.25\textwidth}|}%|p{0.05\textwidth}|p{0.05\textwidth}}
%   \hline
%      Perf Paranoid Setting&Events Access\\\hline%&Kernel Profiling&CPU Profiling \\\hline
%      -1:&Allow use of (almost) all events by all users
%   Ignore mlock limit after perf\_event\_mlock\_kb without CAP\_IPC\_LOCK\\\hline 
% $>= $0: &Disallow raw and ftrace function tracepoint access\\\hline 
% $>= $1: &Disallow CPU event access\\\hline 
% $>= $2: &Disallow kernel profiling\\\hline 
% $>= $4: &Disallow access to HPCs\\\hline 
%   \end{tabular}}
%   \label{tab:perfsetting}
% \end{table}

\subsection{Hardware Performance Counters (HPCs)}
Hardware performance counters are a set of special-purpose registers built-in modern microprocessors to capture the count of hardware-related events. HPCs have been extensively used to predict the power, performance tuning \cite{wang2019a+,makrani2018energy,wang2019survey}, attacks detection \cite{wang2020scarf,wang2020dreal,wang2020hybridg,wang2020phased,wang2021evaluation,sayadi2020recent,wang2020hybrid,wang2021machine,wang2020mitigating,wang2020comprehensive,taram2019fast}, debugging, and energy efficiency of computing systems. It also enhances systems' security by providing microarchitectural information of malware, side-channel attacks, and building detectors based on the events' information \cite{demme2013feasibility,khasawneh2015ensemble,zhang2016cloudradar}. However, recent studies have shown the suitability of such HPCs-based classification for spying on users' behaviors and violating users' privacy \cite{gulmezoglu2017perfweb,naghibijouybari2018rendered}. Such privacy breach attacks gain access to HPCs via Perf tool which is a Linux-based low-level performance monitoring tool and provides considerable functionality and abstraction in the kernel, making the interface straightforward for ordinary users \cite{perfwebpapge,de2010new}. Though Perf is equipped with access control setting, i.e. $perf\_event\_paranoid$, attacks are still able to access HPCs of applications initiated outside kernel space unless $perf\_event\_paranoid$ equals 4. Hence, the HPCs-based fingerprinting attacks still pose significant threats to system security and users' privacy.

\subsection{Machine Learning based Classification}
\subsubsection{Boosting}
Boosting aims to enhance the performance of machine learning algorithms, where the incorrectly classified data from the previous model is employed to implement an an ensemble of models. Compared to Adaboost using exponential loss function, the Logitboost \cite{kamarudin2017logitboost} algorithm uses a binomial log-likelihood that changes the loss function linearly. This attribute makes the target model less sensitive to outliers and noise. To the best of our knowledge, no research to date has investigated the performance of the Logitboost algorithm in the field of webisite fingerprinting attacks. In this work, we applied LogitBoost, %abbreviated as Logit in following section,
as a boosting learning technique on classical machine learning algorithm RandomForest (RandomF) where the boosted model is abbreviated as Logit-RandomF.
\subsubsection{Deep Learning}
% \newline
\textit{Fully Convolutional Neural Network (FCN)} 
A convolutional
layer in a deep neural network learns patterns of local structure
in the input signal and can learn feature representations over a
sequence of input data \cite{lecun1995convolutional}. FCN is based on the convolutional neural network (CNN) technique, where models employ continuous convolution layers to extract time-series features. \textit{Long Short-Term Memory (LSTM)}
For the LSTM \cite{karim2017lstm}, each temporal trace includes a time-ordered sequence of HPCs on which an LSTM network detects temporal patterns that are important for discriminating different websites. One and two layers of LSTM neurons with
various numbers of neurons per layer were explored. Each
node learns a different sequence pattern and the collection of
sequence pattern detectors from all the nodes connected to the
output layer are used to classify each HPC temporal sequence.
\subsubsection{Time-series Classification}
% Time-series datasets exist across various discipline, such as heartbeat in medicine, rates of inflation in economics. And in this work, HPCs capturing tool collects HPCs at microsecond scale and forms a temporal sequence and time-series models are included to compare with proposed ones. 
Time-series classification methods deal with such temporal datasets which is representing them in time-domain format and then calculate the distance as difference between time series \cite{lin2007experiencing}. In this work, three prominent classification algorithms are chosen as representatives to compare with the proposed including the dynamic time warping (DTW) \cite{berndt1994using}, Shaplet \cite{ye2009time}, Bag of Patterns (BOP) \cite{lin2009finding} are selected as representatives. DTW-KNN determines the best alignment that will produce the optimal distance and classifies data according to the calculated distance between time-series sub-sequences. BOP is a structure-based algorithm where a time-series sequence will be transformed into symbolic words while BOP records frequency of each symbol without order information. Shaplet \cite{ye2009time} overcomes the time and space complexity and allows detection for phase-independent shape-based similarity of sub-sequences.
% \subsection{Adversarial Learning}
\subsection{Related Work}
% \subsubsection{Webpage Fingerprinting}
% \subsubsection{Webpage Protection}
Protecting browsing history has emerged recently as a crucial concept to ensure the preservation of users' privacy. %and has drawn a number of researchers. 
 In response, \cite{ylonen2006secure,nithyanand2014glove} are proposed to leverage SSH based protection methods; \cite{ylonen2006secure} encrypts and authenticates messages in one session, and \cite{nithyanand2014glove} adds cover traffic conservatively while maintaining high levels of security. Similarly, Tor project \cite{torproject} is one of the most popular traffic transmission approaches, where messages are not directly routed to the receiver but encrypted and forwarded according to ephemeral paths of an overlay network. Though great progress made by such works, there are still a number of attacks which could bypass such protection mechanisms and extract users' browsing history.    

% \begin{figure}[htb!]
%   \centering
%   \includegraphics[width=0.48\textwidth]{pic/attack model.png}
%   \caption{Proposed attack model}
%   \label{fig:networkattack}
%   % \vspace{-2ex}
% \end{figure}
% \subsubsection{Network Traffic-based Attack}
% There are a number of attacks that capture the network traffic between the web server and client machine by leveraging network namespaces and tcpdump \cite{cai2012touching,rimmer2017automated}. Such works are able to achieve high success rate even if users adopt previous works to encrypted traffic, like SSL/TLS or SSH connections, or anonymity network like Tor \cite{torproject} which routes to different paths. 

% \subsubsection{Operating System-based Attack}
Some recent website fingerprint attacks exploit the clients' machine information when visiting different websites, like memory footprint \cite{jana2012memento}, storage \cite{spreitzer2016exploiting}, etc. To obtain the information, some attacks prepare a malicious website for users to visit, or a local malware to be launched on the target host. For example, \cite{shusterman2020website} launches a Prime+Probe attack to measure cache occupancy through a malicious website. Then, a deep learning method is leveraged to classify websites and recover users browsing history.  Other works such as \cite{jana2012memento} samples the memory footprint of browsers through the procfs file system in Linux. To defend against such attacks, \cite{iqbal2020fingerprinting} proposes ML-based syntactic-semantic approach that detects
browser fingerprinting attacks' behaviors by incorporating both static and dynamic JavaScript analysis. \cite{faizkhademi2015fpguard} proposes to monitor the running Web objects on user's browser and collect fingerprinting related data. Then, it analyzes them and searches for patterns of fingerprinting attempts. Though effective, they only work for attacks deployed through malicious websites. Furthermore, advanced attacks \cite{gulmezoglu2017perfweb} can be deployed in native code and bypass such detection systems. \cite{nikiforakis2015privaricator} randomizes properties, like offsetHeight and plugins, to the JavaScript environment, which generates different fingerprints even for the same website and increases non-determinism for attackers. However, the randomization is complex and can change the visual appearance of websites.

\begin{figure*}[htb!]
    \centering
    \includegraphics[width=0.98\textwidth]{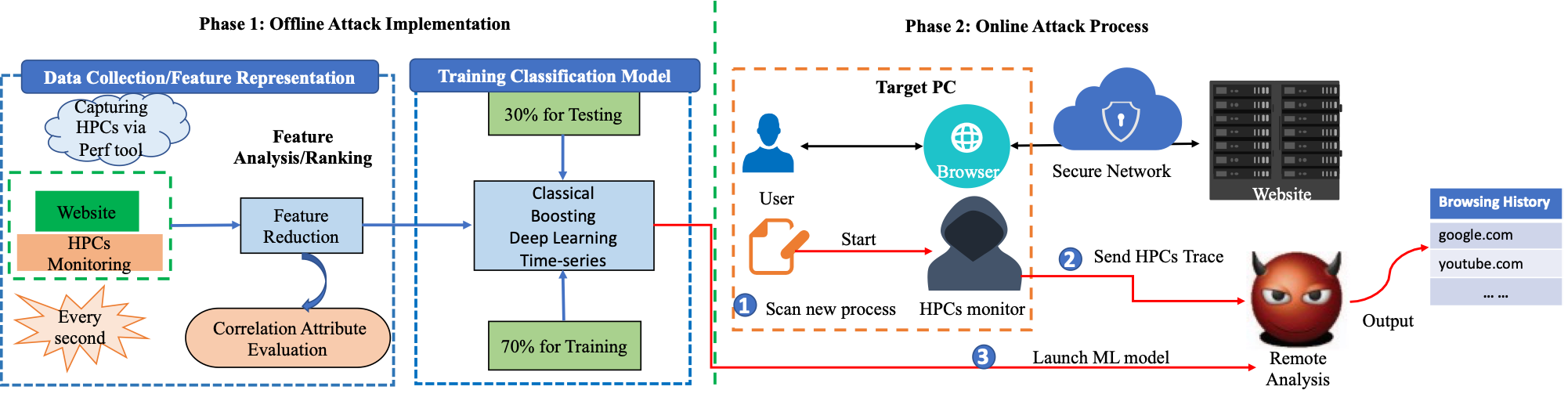}
    \caption{Overview of the proposed {\proj} attack model}
    \label{fig:attackmodel}
    % \vspace{-2ex}
\end{figure*}
% \vspace{-1ex}
\section{Overview of Leaked-Web Attack Model}
\label{sec:model}
This section presents the details of the proposed {\proj} attack model. {\proj} is an accurate and efficient HPC-based attack which fingerprints websites with one local HPCs monitoring unit and a remote machine learning-based trace analyzer as shown in Figure \ref{fig:attackmodel}. During the offline attack implementation phase, HPCs data are collected for each website and the importance (ranking) of HPC features are evaluated. Next, various machine learning algorithms are implemented to find the most effective model using a percentage split training-testing method where 70\% of data (50 traces per website) is assigned to training set and 30\% of data (20 traces per website) is dedicated to testing set. Then, the trained ML model is launched and deployed for online attacking process. For the attacking phase, there are three steps considered in {\proj}: 1) the browser-related process is scanned every second; 2) HPCs monitoring will be initiated once new browser process is found; and 3) the ML classification model is deployed to predict the website's information based on the newly collected HPCs trace.  %Compared to prior works \cite{shusterman2020website,cai2012touching}, {\proj} offers no extra malicious  website containing scripts to obtain cache access patterns, no high sampling rate or large datasize to transfer which causes high performance overhead. %shown in Table \ref{tbl:contribution}. 
\subsection{Threat Model}
As shown in Figure \ref{fig:nativemodel}, for our threat model we consider that the malicious codes reside in the host machine, initiate HPCs collection, and send them to a remote attacker. The malicious codes can be launched by inserting it into benign applications or downloading accidentally. With this model, the website fingerprinting attacks can be activated as long as the malicious codes are installed already and do not need users to open certain malicious websites. Furthermore, $perf\_event\_paranoid$ is set less than 4, giving access to reading HPCs from registers. The attacker is able to obtain the potential websites users might open at Alex top site \cite{analitic2020top}. 
% In this attack model, the assumption is that the malicious code is resided in the host machine already, which can be done by inserting it into benign applications or downloading accidentally. With this model, attacks can be activated as long as the malicious codes are installed already and do not need users to open certain websites.
\subsection{Experimental Setup}
In this work, all experiments are conducted on an Intel i5-3470 desktop with 4 cores and 8GB DRAM, three-level cache system.
In this on-chip cache memory subsystem, while L1 and L2 caches are exclusively separated, the L3 cache memory is inclusive and shared among all cores. In addition, the operating system is Ubuntu 20.0.4 LST with Linux kernel 5.8.0. The proposed HPC-based attack is implemented on a widely used web browsers, Firefox. 

% \section{Data Collection}
\begin{figure}[htb!]
    \centering
    \includegraphics[width=0.48\textwidth]{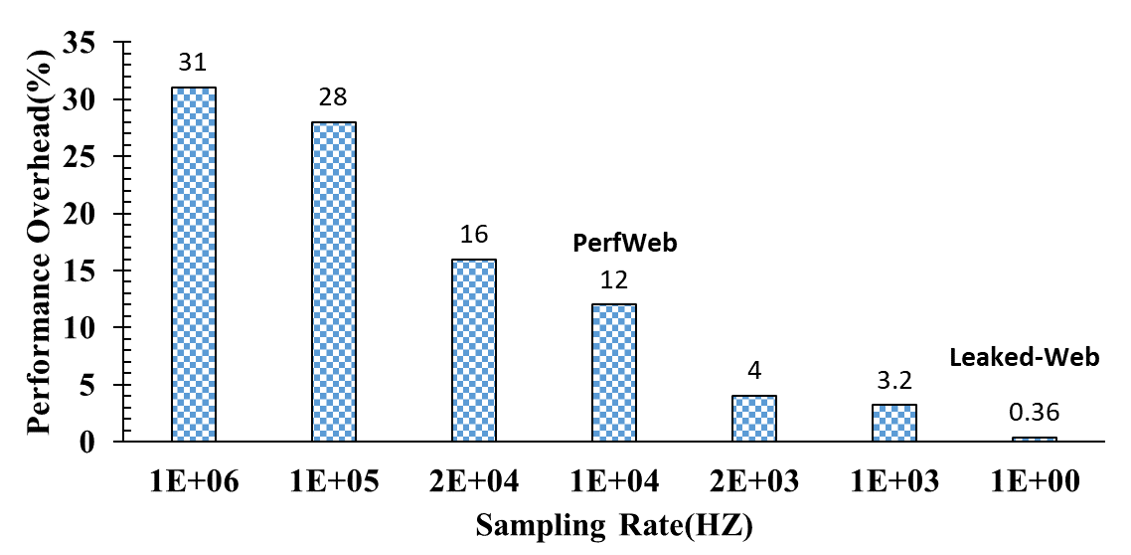}
    \caption{Performance overhead with various sampling rates (HZ)}
    \label{fig:overhead}
    % \vspace{-2ex}
\end{figure}
% \begin{figure}[htb!]
%     \centering
%     \includegraphics[width=0.48\textwidth]{pic/size.png}
%     \caption{Samples and datasize per trace comparison across various works}
%     \label{fig:datasize}
%     % \vspace{-2ex}
% \end{figure}
\subsection{Hardware Performance Events Monitoring}
In this work, we use Perf \cite{perfwebpapge} to measure the hardware-related features, and memory and processor's low-level behavior. Perf is a profiling and performance analysis tool that can help to track the hardware performance counters. 
As introduced in Section \ref{sec:bg}, any value less than 4 for $/proc/sys/kernel/perf\_event\_paranoid$ gives users access to the HPCs-based profiling of website process. %To capture the newly opened website, the attacker scans running process every second and starts monitoring with the process id once detected. 
% The monitoring command is like:
% \begin{verbatim} 
% perf stat -e event -I 1000 -o out_file -p pid.
% \end{verbatim} 
Additionally, %as mentioned earlier %in Section \ref{sec:model}, 
we %compare
examine the performance overhead and sample size caused by HPCs monitoring in Figure \ref{fig:overhead}. As depicted, the x-axis represents applied the sampling rate ranging from \(1^6\) to \(1^0\), the primary y-axis %represents
denotes the execution time of victim applications, and the second y-axis represents performance overhead under different sampling rate. Moreover, execution time under no HPCs monitoring is used to obtain %get
the performance overhead percentage. It is observed that generally, the smaller the %of
sampling rate is, the larger the performance overhead is. For instance, when the monitoring scale is \(1^6\), the performance overhead is at its highest value %and 
reaching to 30\%. Hence, to make the influence of sampling rate on system performance and the proposed attack less noticeable, we choose \(1^0\) for the HPCs monitoring in {\proj}. % To further boost prediction accuracy without increasing the number of HPCs or samples, we collect the 16 HPCs as listed in Table \ref{tbl:hpcs} for evaluating the importance and selecting the most effective ones during training phase. During the testing, only the selected HPCs are collected to eliminate the need for multiple profiling.

%  \vspace{-2ex}
\begin{table}[!htb]
  \centering
      \caption{The collected HPC features and their ranking}
  \scalebox{1.0}{%
  \begin{tabular}{|l|l|l|l|}
    \hline
    % \multicolumn{4}{c}{\textbf{HPCs Importance}}\\\hline
    \textbf{Rank}&\textbf{HPC} &\textbf{Rank}&\textbf{HPC}\\\hline
1&cache-misses&9&branch-instructions\\\hline
2&node-loads&10&iTLB-loads\\\hline
3&branch-misses&11&iTLB-load-misses\\\hline
4&branch-load-misses&12&dTLB-store-misses\\\hline
5&LLC-store-misses&13&dTLB-load-misses\\\hline
6&branch-loads&14&dTLB-stores\\\hline
7&L1-dcache-stores&15&node-stores\\\hline
8&L1-icache-load-misses&16&L1-dcache-load-misses\\\hline
    \end{tabular}}
    %   \vspace{-2em}
  \label{tbl:hpcs}
\end{table}
%  \vspace{-3ex}
% \begin{table}[!htb]
%   \centering
%       \caption{The collected HPC features and their ranking}
%   \scalebox{0.85}{%
%   \begin{tabular}{l|l}
%     \hline
%     % \multicolumn{4}{c}{\textbf{HPCs Importance}}\\\hline
%     \textbf{HPC Name} &\textbf{HPC Name}\\\hline
% L3 MISSES&BRANCHES\_MISPREDICTED\\\hline
% L2\_HIT&ITLB\_MISSES\\\hline
% L1\_MISS&DTLB\_LOAD\_MISSES\\\hline
%  L1 HIT& L3 HIT \\\hline 
% INST\_RETIRED.ANY&ALL\_BRANCHES\_RETIRED\\\hline
% UOPS\_RETIRED.ALL&DTLB\_STORE\_MISSES\\\hline
% L2\_MISS&BR\_TAKEN\_CONDITIONAL\\\hline
% BR\_NONTAKEN\_CONDITIONAL &TAKEN\_INDIRECT\_NEAR\_CALL \\\hline
%     \end{tabular}}
%     %   \vspace{-2em}
%   \label{tbl:hpcs}
% \end{table}
% \vspace{3ex}
\subsection{Database Description}
\label{sec:database}
We %choose
select the top 30 websites from Alexa Top Sites \cite{analitic2020top}. Similar to previous works no traffic modeling is applied in our database implementation. %This work follows similar method of previous work where no traffic modeling is applied. 
For the purpose of thorough analysis, this work considers both Closed World and Open World datasets as described below:

\subsubsection{Closed World Dataset}The closed world dataset means that each website is sensitive and exists in training dataset. The proposed attack model considers distinguishing a relatively small list of websites (30) and each websites has 50 traces for training a classification model and 20 traces for testing.

\subsubsection{Open World Dataset} %Due to the criticism of the closed world assumption for the requirement 
Besides the sensitive websites mentioned in the closed world dataset, open world dataset also contains a large set of non-sensitive web pages, all of which the attacker is expected to generally label as “non-sensitive” \cite{shusterman2020website}.
% Given that the closed world dataset requires a complete knowledge of %for the knowledge of the complete 
% website list that the users will visit, in this work we have also collected an %also prepares an 
% open world dataset. In the open world dataset, the attack only access the knowledge of a set of sensitive websites %that will 
% to be monitored and further has the ability to classify them correctly. At the same time, there are a large set of non-sensitive websites which the attacker should label as "no-sensitive". 
For the open world dataset, we add additional 500 traces in which each of them %is collected from
represents the behavior of a single unique website. 
% \subsubsection{Target  website}

% \subsection{Hardware Settings}
%Here, the details of the configurations deployed and SCAs used in this work.
\begin{figure}[htb!]
    \centering
    \includegraphics[width=0.48\textwidth]{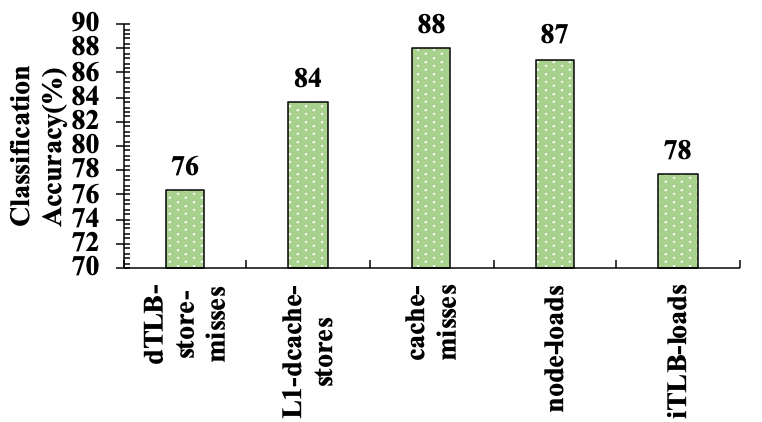}
    \caption{The average classification accuracy under Firefox for each HPC}
    \label{fig:hpc}
    % \vspace{-2ex}
\end{figure}

%  \vspace{-2ex}

\subsection{HPC Events Importance Evaluation}
\label{sec:hpceval}
Figure \ref{fig:hpc} compares the accuracy of Logit-RandomF-based classifier for websites classification using different HPC events (due to space limitations only 5 events are reported). %(only five events are listed d brief reason). 
As can be seen in this figure, changing the HPC %only can 
could result in over 10\% accuracy loss when the same ML classifier is applied. Furthermore, given that there exists a limited number of HPC registers physically available on modern microprocessors' chips that can be accessed %collected %at one time
simultaneously \cite{intel2007ia},
it is necessary to identify the most important HPCs for classifying the %victim
 websites. %For HPCs reduction, 
To select the most prominent HPC features we employ Correlation Attribute Evaluation ($CorrelationAttributeEval$ in Weka \cite{hall2009weka}) with its default settings to calculate the Pearson correlation between attributes (HPC features) and classes (websites). %And then the sum score of each HPC features (min, max, stdev, and sum in this work) will be calculated and HPCs will be ranked according to sum score as shown in Table \ref{tbl:hpcimportance}. %\textcolor{red}{After reduction, only selected HPCs will be monitored during testing or validating phase.}
%\textcolor{blue}{
Correlation attribute evaluation algorithm calculates the Pearson 
correlation coefficient 
between each attribute and class, as given below:
% \vspace{-1em}
\begin{equation}\label{eq:corr}
\centering
% \vspace{-1ex}
\rho(i)=\frac{cov(Z_i,C)}{\sqrt{var(Z\_i)\ var(C)}} \quad i=1,...,16
\end{equation}
where $\rho$ is the Pearson correlation coefficient. $Z_i$ is the input dataset 
of event $i$ ($i=1,\dots,16$). $C$ is the output dataset containing labels, % different classes, 
i.e., websites, like "google.com", "youtube.com" and etc. in our case. 
The $cov(Z_i, C)$ measures the covariance between input data and output 
data. The $var(Z_i)$ and $var(C)$ measure variance of both input and 
output datasets, respectively. Next, the HPCs will be ranked according as shown in Table \ref{tbl:hpcs} and this work chooses the top 4 HPCs for classification.
\subsection{Machine Learning Classifiers}
In the proposed {\proj} attack, supervised learning is used to model the  website fingerprinting attack. The ML implementation stage consists of a building step (training) and attack step (testing). In the building step, multiple traces (50) from each website are collected and %and the traces are
labeled. The labeled dataset is used to train%s 
various types of ML classifier including classical machine learning, classical machine learning with boosting, time-series, or deep learning models. The rationale for choosing these machine
learning models is that they are from different branches
of ML including classical model (RandomForest, LogitBoost RandomForest), deep learning models (FCN, LSTM), time-series models (DTW, BOP, Shapelet) techniques covering a diverse range of learning
algorithms that support our comprehensive analysis and experiments. For the attacking phase, the proposed attack model receives unlabeled traces in which each of the trace is corresponding to a user's website visit and the trained classifier %gives
outputs the prediction results. Each website has 20 traces for testing and the accuracy is calculated by comparing correctly classified %predicted
labels and actual samples.
\section{Experimental Results and Analysis}
\label{sec:eval}
In this section, we comprehensively evaluate the effectiveness of the proposed {\proj} attack model in terms of classification accuracy and F-measure (F-score) analysis with different ML models, number of HPCs features, monitoring duration. Such analysis gives insight into the cost of HPCs-based fingerprinting attacks and further indicate the requirements of protection approaches.  %in terms of classification accuracy, F-measure under Firefox and Chrome browsers.  
% Additionally, we demonstrate the training and testing latency of the proposed {\proj} and previous work \textit{Perf-Web} \cite{gulmezoglu2017perfweb}.

\begin{figure}[!htb]
    \centering
    \includegraphics[width=0.48\textwidth]{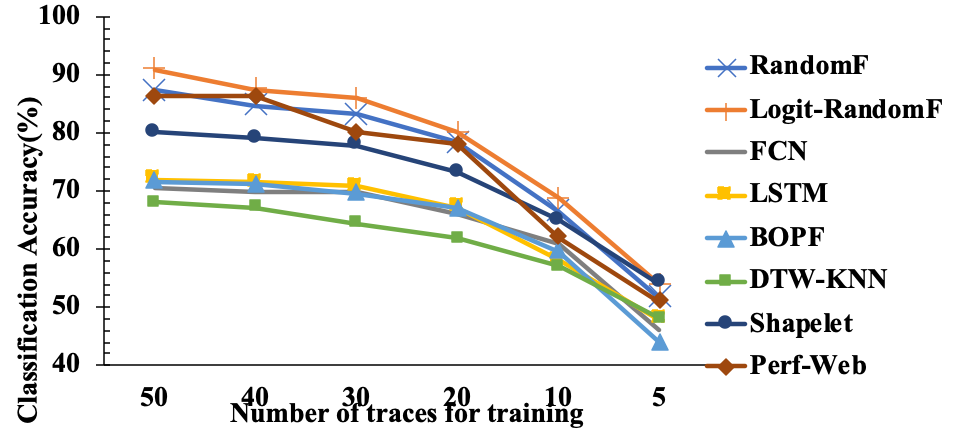}
    \caption{Classification accuracy with various classification algorithms with 4 HPCs}
    \label{fig:cmpfirefox}
    % \vspace{-2ex}
\end{figure}
\begin{figure}[htb!]
    \centering
    \includegraphics[width=0.48\textwidth]{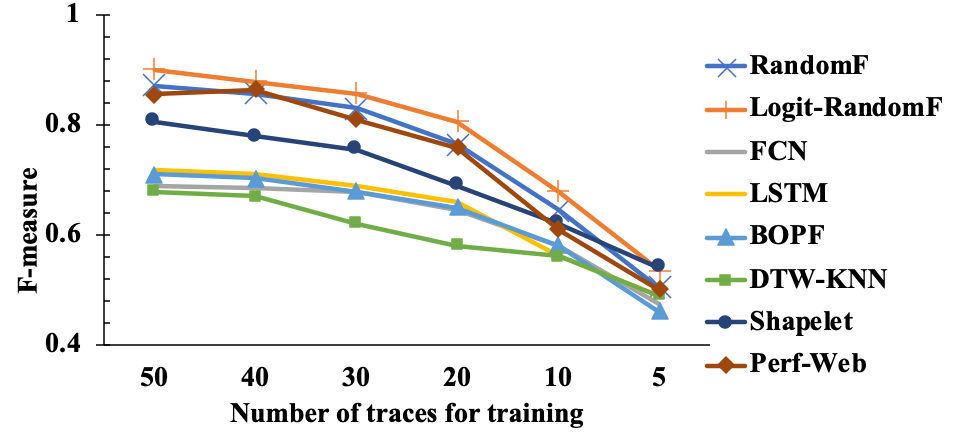}
    \caption{F-measure with various classification algorithms 4 HPCs}
    \label{fig:cmpchrome}
    % \vspace{-2ex}
\end{figure}
% \subsection{HPCs Effectiveness Evaluation}
% Due to the limited of HPCs can be collected simultaneously, it is crucial to identify the importance of each HPC in terms of classifying websites accurately and select the effective ones for building more effective HPCs. As shown in Figure \ref{fig:hpc}, all 16 HPCs are categorized into XX types: cache-related, memory-related, storage-related, branch-related. The classification accuracy of each HPC with the proposed LogitBoost-RandomF is depicted and compared.

% \subsection{Closed World Results}

\subsection{ML Classification Models Comparison}
As introduced in Section \ref{sec:database}, %there are 
30 websites selected from the Alexa ranking are executed on our target system and each website has 50 traces for training and 20 traces for testing. Various ML classification models from classic, boosting, time-series and deep learning methods are investigated.
As shown in Figure \ref{fig:cmpfirefox} and Figure \ref{fig:cmpchrome}, X-axis represents the number of traces for training in selected classification models from four ML types (classic, boosting, time-series, and deep learning methods) and Y-axis represents the classification accuracy and F-measure respectively for closed world. F-measure is interpreted as a
weighted average of the precision (p) and recall (r) which
is formulated as \(\frac{2 \times (p\times r)}{p+r}\). The precision is the proportion
of the sum of true positives versus the sum of positive
instances and the recall is the proportion of instances that
are predicted positive of all the instances that are positive. F-measure is a comprehensive evaluation metric %over accuracy (percentage of correctly classified samples) 
since it
takes both the precision and the recall into consideration. More
importantly, F-measure is also resilient to class imbalance in
the dataset which is the case in our experiments. 

As observed from the results, generally, reducing number of traces in training phase reduces both classification accuracy and F-measure values. %Especially 
This observation becomes more noticeable as we reduce the traces to lower than 20 where the accuracy becomes below to 80\% for all the applied ML classification models. Another interesting observation %noticeable result
is that Logit-RandomF classifier outperforms the previous work \textit{Perf-Web} \cite{gulmezoglu2017perfweb} for most of the training sizes (except when the number of traces for training drops to 5). When training traces is 50, Logit-RandomF classifier achieves the highest classification accuracy and F-measure,  91\% and 0.901 respectively. As seen, the accuracy is improved in {\proj} by around 5\% from 86\% of previous work \cite{gulmezoglu2017perfweb}. Another observation is that Shaplet-based classification model has shown to be more effective than the rest of two time-series classification models.

%Additionally, F-measure of the proposed Logit-RandomF is  which is also the highest when the number of training traces is 50.  % As shown in Figure \ref{fig:cmpfirefox} and Figure \ref{fig:cmpchrome}, X-axis represents the the selected classification models from three types, ie., classic, time-series and deep learning methods; Y-axis represents the classification accuracy for close world. It can be observed that LogitBoost-RandomF outperforms the best, having 90.1\% and 91.3\% improving by around 4\% from 86.4\% of previous work. % Though previous works present that deep learning models perform the best, 
\begin{figure}[htb!]
    \centering
    \includegraphics[width=0.48\textwidth]{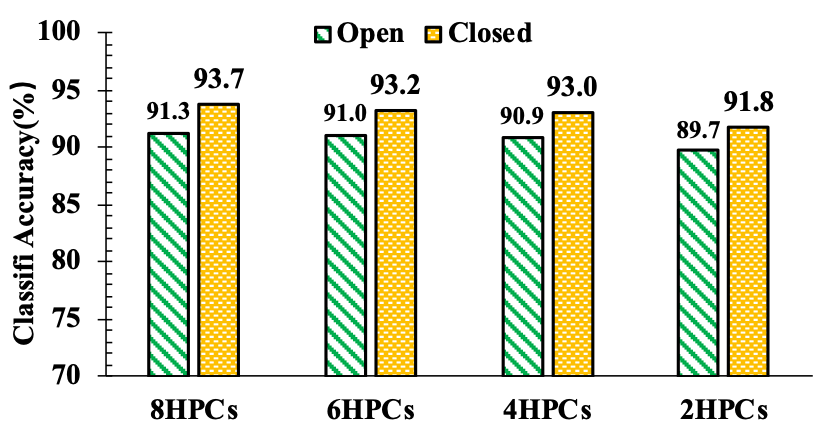}
    \caption{Classification accuracy with various number of HPCs features with Logit-RandomF for closed and open world dataset}
    \label{fig:hpcs}
    % \vspace{-2ex}
\end{figure}
\begin{figure}[htb!]
    \centering
    \includegraphics[width=0.48\textwidth]{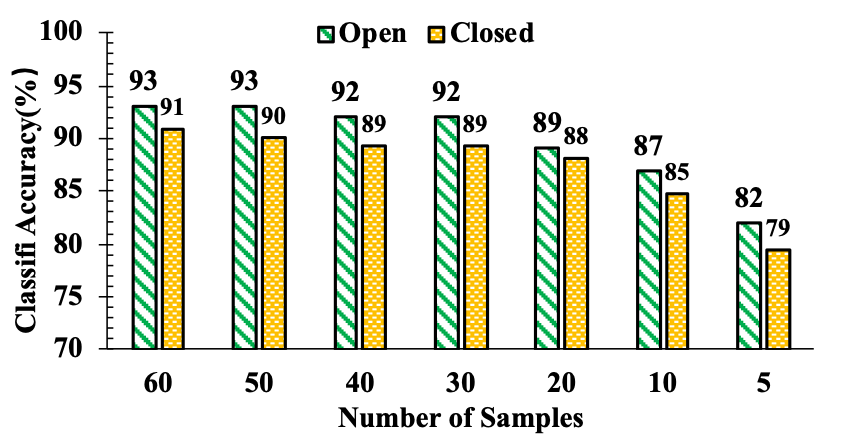}
    \caption{Classification accuracy with various number of samples per trace for closed and open world dataset}
    \label{fig:samplesfox}
    % \vspace{-2ex}
\end{figure}
% \vspace{-2ex}
% \begin{figure}[htb!]
%     \centering
%     \includegraphics[width=0.45\textwidth]{pic/open.png}
%     \caption{Classification accuracy with various number of samples per trace under open world dataset}
%     \label{fig:open}
%     % \vspace{-3ex}
% \end{figure}
%\vspace{-2ex}
\subsection{The number of HPC Features}
As our analysis showed that the Logit-RandomF model performs best among all experimented ML models in {\proj}. Hence, we explore the classification accuracy of Logit-RandomF model under various number of HPCs features and samples, which gives further insights into the potentially vulnerable architectures and duration of attacks. Such analysis is important for evaluating the effectiveness and complexity of future protection mechansims against such privacy violation attacks. This section primarily examines the accuracy of HPC-based website fingerprinting attacks when the number of HPCs changes from 8 to 2, indicating the leakage potential under other architectures with less or more available HPCs registers. As shown in Figure \ref{fig:hpcs}, the accuracy remains above 89\% and 91\% for both open and closed dataset when the number of HPCs features reduces from 8 to 2. This indicates that such HPC-based fingerprinting attacks can be effective in accurately inferring users' browsing history at run-time in processor architectures with varying number of HPC registers even with limited available resources (only 2 HPC registers). 
 
\subsection{Monitoring Duration}
%\subsubsection{Closed World Results}
% \noindent
% \textit{- Closed World Results}
%From above Section, it is safe to 
In this Section, we further investigate the %time duration for monitoring HPCs
HPCs monitoring duration in order to maintain a high classification performance.
Since the number of samples per trace directly decides the duration for data collection when the attack is launched, using less samples %meaning
indicates that the %proposed
attack can be applied even when users visit a website within less than 1 minute. As shown in Figure \ref{fig:samplesfox}, X-axis represents the number of samples ranging from 60 to 5 which means monitoring website for 60 second to 5 second.  It can be observed that when reducing the number of samples from 60 to 20, the classification accuracy for closed and open world dataset has slight decrease from 90\% to 88\%, and 93\% to 89\% for closed and open world dataset. However, further reduction from 20 samples to 10 samples and then to 5 samples per trace causes more significant reduction, from 88\% to 84\% and then to 79\%. Though the noticeable decrease of classification accuracy with only 5 samples, they accuracy still remains around 80\%, indicating the capability of inferring websites of the attacker within 5 seconds.

\section{Conclusion}
\label{sec:con}
Website fingerprinting attacks have emerged recently  through stealing users’ online behaviors and access patterns, to induce users' personal and sensitive information. In  hardware-assisted website fingerprinting attacks, when users open a website in a web browser, they leave a distinct pattern on the underlying hardware that is reflected in the microarchitectural state of the processor running the browser. While Hardware Performance Counters (HPCs) are widely used for performance tuning, application profiling, malware detection, etc., this work presents {\proj}, a fast and unified HPC-based attack model that collects microarchitectural HPC samples by using Perf tool under Linux and trains accurate and efficient machine learning classifiers with the HPCs' traces. Compared to prior works, {\proj} demands significantly lower network traffic per website visit and achieves up to 91\% classification accuracy outperforming the state-of-the-art attack by nearly 5\%. We also explored that the accuracy under different number of HPCs features is around 90\% even with 2 HPC features, indicating that the proposed HPCs-based attack is effective to be adopted in other modern processor architectures (e.g., ARM) with less available HPC registers. %Additionally, the monitoring duration of the proposed attack can be reduced to only 5 seconds but still yields around 80\% accuracy.
Furthermore, our proposed attack obtains a trivial performance overhead (less than 1\%) which is more than 12\% lower than the existing HPC-based attacks. 

% trivial performance overhead and demands much fewer network traffic per webpage visit. At the same time, classification accuracy improves from 86\% to 91\% with the usage of LogitBoost.
\bibliographystyle{acm}
\bibliography{acmart.bib}
\end{document}